\documentclass[useAMS]{mn2e}
\usepackage{graphics,graphicx}
\usepackage{lscape}
\usepackage{rotating}
\usepackage{amssymb}

\def\msun{\hbox{M$_\odot$}}

\def\t4{\hbox{t$_{\rm 4}$}}

\def\cm3{\hbox{cm$^{-3}$}}

\input psfig.sty
\input epsf.sty
\usepackage{graphicx}
\usepackage{epsfig}
\title[eMSTOs Are Not Caused By Age Spreads]
{The Morphology of the Sub-Giant Branch and Red Clump Reveal No Sign of Age Spreads in Intermediate Age Clusters}
\author[Bastian \& Niederhofer] {N. Bastian$^{1}$ \& F. Niederhofer$^{2,3}$ \\
$^{1}$ Astrophysics Research Institute, Liverpool John Moores University, 146 Brownlow Hill, Liverpool L3 5RF, UK\\
$^{2}$ European Southern Observatory, Karl-Schwarzschild-Stra{\ss}e 2, D-85748 Garching bei M\"unchen, Germany \\
$^{3}$ Universit\"ats-Sternwarte M\"unchen, Scheinerstra{\ss}e 1, D-81679 M\"unchen, Germany \\
}
\date{Accepted. Received; in original form}
\pagerange{\pageref{firstpage}--\pageref{lastpage}}
\pubyear{2014}
\begin{document}
\maketitle
\label{firstpage}
\begin{abstract}

A recent surprise in stellar cluster research, made possible through the precision of Hubble Space Telescope photometry, was that some intermediate age ($1-2$~Gyr) clusters in the Large and Small Magellanic Clouds have main sequence turn-off (MSTO) widths that are significantly broader than would be expected for a simple stellar population (SSP).  One interpretation of these extended MSTOs (eMSTOs) is that age spreads of the order of $\sim500$~Myr exist within the clusters, radically redefining our view of stellar clusters, which are traditionally thought of as single age, single metallicity stellar populations.  Here we test this interpretation by studying other regions of the CMD that should also be affected by such large age spreads, namely the width of the sub-giant branch (SGB) and the red clump (RC).  We study two massive clusters in the LMC that display the eMSTO phenomenon (NGC~1806 \& NGC~1846) and show that both have SGB and RC morphologies that are in conflict with expectations if large age spreads exist within the clusters.  We conclude that the SGB and RC widths are inconsistent with extended star-formation histories within these clusters, hence age spreads are not likely to be the cause of the eMSTO phenomenon.  Our results are in agreement with recent studies that also have cast doubt on whether large age spreads can exist in massive clusters; namely the failure to find age spreads in young massive clusters, a lack of gas/dust detected within massive clusters, and homogeneous abundances within clusters that exhibit the eMSTO phenomenon.

\end{abstract}

\section{Introduction}
\label{sec:intro}

A major open topic in stellar cluster research is whether clusters can host multiple star-formation events or extended star-formation epochs.  Young massive clusters (YMCs) have stellar populations that are well reproduced as single burst events, with upper limits on potential age spreads of $1-3$~Myr (cf. Longmore et al.~2014).  Globular clusters, on the other hand, host chemical abundance anomalies within their stellar populations (e.g., Gratton et al.~2012), and while age spreads cannot be discerned at these old ages, some scenarios to explain the anomalies invoke multiple star-forming epochs, spanning tens to hundreds of Myr (e.g., D'Ercole et al.~2008; Conroy \& Spergel~2011), although other scenarios have little or no age spreads (e.g., Krause et al.~2013; Bastian et al.~2013b).

Mackey \& Broby Nielsen~(2007) constructed a colour-magnitude diagram (CMD) for NGC~1846, an intermediate age ($1-2$~Gyr) massive ($\sim10^5$\msun) cluster in the LMC, and found that the main sequence turn-off region was not well described by a simple stellar population (SSP), but rather was extended.  The authors interpreted this spread as a spread in age of $\sim300$~Myr and postulated that NGC~1846 was the result of a rare cluster merger, where the two clusters had a $\sim300$~Myr age difference.  However, subsequent work has shown that the majority of intermediate age ($1-2$~Gyr) clusters in the LMC and SMC display this extended main-sequence turn-off (eMSTO - e.g., Mackey et al.~2008; Milone et al.~2009; Goudfrooij et al.~2011a,b; Keller et al.~2012; Girardi et al.~2013), ruling out rare events like unequal age cluster mergers.  If interpreted as being caused by age spreads, these results suggest, rather, that star-formation can proceed for $300-600$~Myr within massive clusters, and that age spreads of this magnitude should be a common feature in massive clusters.

Such age spreads have been searched for in younger massive clusters, where such spreads should be readily apparent.  Larsen et al.~(2011) studied the resolved CMDs of six massive ($10^5 - 10^6$\msun) clusters in nearby galaxies, and did not find evidence for age spreads of more than $20-30$~Myr (and being consistent with an SSP in most cases).  Bastian \& Silva-Villa~(2013) looked at the CMDs of NGC~1866 and 1856, two clusters with masses of $\sim10^5$\msun, and ages of 180 and 280~Myr, respectively.  No age spreads were found, with upper limits of any potential spreads of $\sim30$~Myr.  Niederhofer et al.~(2015) extended the Bastian \& Silva-Villa survey to an additional eight LMC young massive clusters, including NGC~1850, a $\sim2\times10^5$, 100~Myr cluster, and also did not find any evidence for age spreads.  It is important to note that the YMCs studied so far have current masses (and escape velocities) well in excess of the intermediate age clusters that show eMSTOs. Hence, unless extreme cluster mass loss is invoked, the YMCs have the same properties (or even higher mass and escape velocities) as the intermediate age clusters that display the eMSTO features did when they were at a similar age.  Finally, Bastian et al.~(2013a) examined the integrated spectra (or resolved CMDs) of $\sim130$ YMCs with ages between $10-1000$~Myr and masses of $10^4 - 10^8$\msun, looking for evidence of ongoing star-formation, and none were found.  These studies have called into question the interpretation of the eMSTO feature being due to age spreads.

Since age spreads are not seen in YMCs, alternative explanations for the eMSTO have been put forward.  Bastian \& de Mink~(2009) suggested that stellar rotation may cause the eMSTO feature, as stellar rotation effects the structure of the star and the inclination angle of the star relative to the observer will change the effective temperature, hence observed colour. However, Girardi et al.~(2011) calculated isochrones for a modestly rotating star and found that the extended lifetime of the star, due to rotation, may cancel the stellar structure effects, resulting in a non-extended MSTO.  Platais et al.~(2012) found that in the intermediate age ($\sim1.3$~Gyr) Galactic cluster Trumpler~20, that modestly rotating stars were blue shifted from the nominal main sequence, in agreement with the Girardi et al.~(2011) predictions, suggesting that stellar rotation may not be the cause of the eMSTO phenomenon.  However, no rapid rotating stars were present in the cluster, and further studies are needed to confirm this as a general result.

Yang et al.~(2013) used the Dartmouth stellar evolutionary code to investigate these contradictory results, and found that the mixing efficiency caused by rotation is a key parameter, and that in fact stellar rotation may cause eMSTOs, depending on the specific choice of the rotational mixing parameter.  Milone et al.~(2013) found that while the young ($\sim150$~Myr) cluster NGC~1844 in the LMC did not have an age spread, its CMD does display features that are not well described by an SSP model, namely a broadened main sequence at similar magnitudes as the eMSTO in the intermediate age clusters.  Hence, as CMDs are studied in more detail, new features are still being discovered.

Hence, the cause of the eMSTO is still under debate, with some studies still advocating age spreads as the best explanation (e.g., Correnti et al.~2014)\footnote{Although note that these scenarios use ``correction factors" to translate the observed present day masses and radii of the clusters to the initial values. These factors invoke extreme cluster mass loss, causing the intermediate age clusters to have been significantly more massive in the past.  The models that are used for this (D'Ercole et al.~2008) were made for inner Milky Way tidal fields, i.e. they were not meant for the LMC/SMC and hence would predict significantly smaller mass loss in these smaller galaxies. }. One way to test the above scenarios is to look for other features in the CMD that may also reflect age spreads.  Girardi et al.~(2009) introduced the red clump as such a test, showing that the red clump in NGC~419 is more elongated than would be expected in a simple stellar population.   Interestingly, Trumpler 20 also displays an extended or ``dual red clump" although it does not show any evidence for an age spread within it (Platais et al.~2012).  Li et al.~(2014a) found, on the other hand, that the younger eMSTO cluster NGC~1831, has a compact red-clump which is inconsistent with the presence of a significant age spread within the cluster.  It is currently unclear what the role of rotation will be on the red clump morphology.  Extra-mixing from rotation would likely give larger core masses, which may result in an extended red clump, however this effect may be compensated by a longer main sequence lifetime.  Detailed modelling is necessary to address this.

Another feature of the CMD that is sensitive to age spreads is the width of the sub-giant branch (e.g., Marino et al.~2012).  Li et al.~(2014b) have shown that the SGB width in NGC~1651, an intermediate age LMC cluster that displays the eMSTO phenomenon, is consistent with a simple stellar population (i.e., with no age spread). The authors place an upper limit of 80~Myr on any age spread within the cluster, significantly less than the age spread of 450~Myr inferred from the eMSTO.  While the SGB width does suffer from lower numbers of stars than the MSTO or red clump regions of the CMDs, a number of clusters contain enough stars to place strict limits on any age spreads present within the cluster.

Here we use the same technique as Li et al.~(2014b), studying the width of the SGB in two intermediate age LMC clusters that display the eMSTO phenomenon, namely, NGC~1806 \& NGC~1846.  In \S~\ref{sec:obs} we present the observations, while in \S~\ref{sec:results} we analyse the CMDs of the two clusters, testing the hypothesis of significant age spreads within the clusters.  Finally, in \S~\ref{sec:conclusions} we discuss our results and present our conclusions.

\section{Observations and Models}
\label{sec:obs}

For the present work we use optical Hubble Space Telescope (HST) imaging taken with the Advanced Camera for Surveys (ACS) Wide Field Camera (WFC) from programmes GO-9891 (PI Gilmore) and GO-10595 (PI Goudfrooij).  The observations, reduction, and photometry are all described in detail in Milone et al.~(2009).  We use the field-star subtracted (based on a suitable nearby reference field) catalogues for our analysis, and refer the interested reader to Milone et al.~(2009) for details on this process.  We have corrected the observed photometry for extinction, using values of $A_V = 0.05$ and $0.08$  and distance moduli of 18.5 and 18.42 for NGC~1806 and 1846 (Goudfrooij et al.~2011b),  respectively.  The values were taken from Goudfrooij et al.~(2011b), although independent analysis using the present datasets found good agreement.

For our analysis we adopt the stellar isochrones of Marigo et al.~(2008), with Z=0.008 and Y=0.25.  However, we note that the conclusions of this paper are unchanged if we would have adopted the BaSTI (Pietrinferni et al.~2004) or the Dartmouth (Dotter et al.~2008) isochrones with similar parameters. While the absolute ages of the clusters do depend on the model adopted, the relations between the MSTO, SGB and RC positions remain unchanged, i.e. if an age spread is present it will show up in the same way in each of the different portions of the CMD. 

\section{Results}
\label{sec:results}

\subsection{NGC~1806}
\label{sec:ngc1806}

Fig.~\ref{fig:ngc1806_cmd} shows the CMD of NGC~1806, centred on the MSTO, SGB, and red clump portions of the distribution.  (Blue) lines show isochrones of age 1.41 (top), 1.58 (middle) and 1.86~Gyr (bottom), i.e. the range required to explain the observed spread in the MSTO region.  Some of the spread observed in the MSTO region is due to binaries and photometric errors, hence the isochrones used to bracket the eMSTO are upper limits to any actual spread within the clusters.  It should be noted that different authors have derived somewhat conflicting age spreads, with Goudfrooij et al. (2011b) suggesting age spreads of $\sim500$~Myr for NGC~1806 and Milone et al.~(2009) suggesting a lower value of $\sim200$~Myr. 
We identify candidate SGB branch stars between $1.0 \leq F435W - F814W \leq1.6$ as (red) diamonds. 

As can already be seen from the figure, the SGB stars cluster around the youngest isochrone needed to explain the eMSTO feature, and do not span the full range expected if large age spreads were present in the cluster.

\subsubsection{Photometric Errors}
The average photometric errors were derived from the observed main sequence distribution below the MSTO.  To do this we ``verticalised" the main sequence between $21 \le m_{\rm F435W}  \le 23.5$.  We then made histograms of the stars in  $(F435W - F814W)_{\rm vertical}$ in 0.25~magnitude bins, and fit gaussians to the distribution.  While there was clearly a tail of stars to the red (due to binaries) the core of each distribution was well described by a gaussian with dispersion of $0.028$~mag.  We took this to be the typical error in colour and correspondingly an error of $0.02$~mag in brightness.

\subsubsection{Inferred Age Spreads from the MSTO and Synthetic Cluster CMDs}
\label{sec:msto_cut}

\begin{figure}
\centering
\includegraphics[width=8cm]{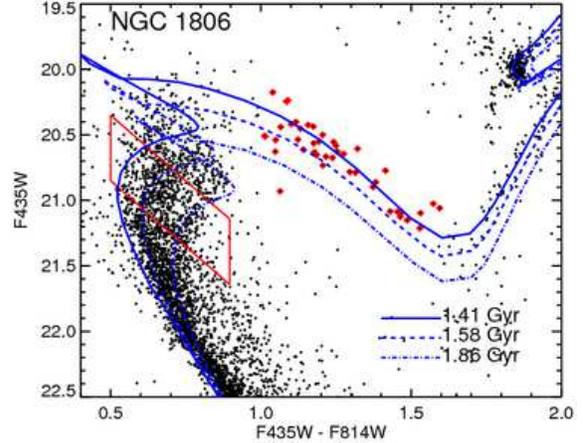}

\caption{The CMD of NGC~1806 along with isochrones of ages 1.41, 1.58, and 1.86~Gyr, which span the width of the eMSTO.  Stars identified as part of the SGB are highlighted in red. Note that the SGB stars do not span the same width in age as the eMSTO.  The red box denotes the cut across the MSTO where the extended star-formation history is inferred from.}
\label{fig:ngc1806_cmd}
\end{figure}

\begin{figure}
\centering
\includegraphics[width=8cm]{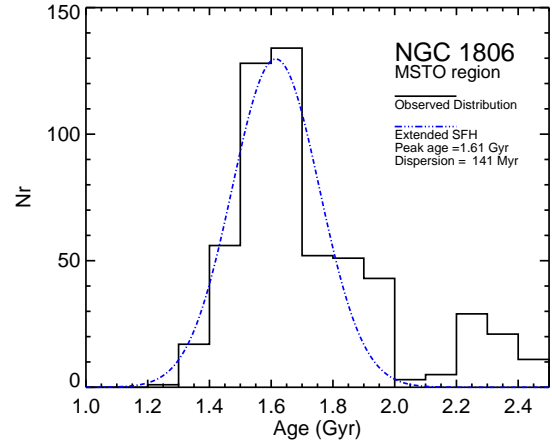}

\caption{The extended star-formation history inferred from the cut across the MSTO.  The (blue) dash-dotted line shows the best fit gaussian to the distribution, with the parameters listed in the panel.}
\label{fig:ngc1806_age_dist_msto}
\end{figure}

We begin by estimating the age spread present as inferred by the extended MSTO.  To do this, we take a slice through the data (shown as the solid red box in Fig.~\ref{fig:ngc1806_cmd}), and find the best fitting isochrone for each observed star. This is similar to what was done in Goudfrooij et al.~(2011b; 2014), although we note that we have not attempted to include the effects of binaries in this simple calculation.  The resulting age distribution is shown in Fig.~\ref{fig:ngc1806_age_dist_msto}.  We then fit a gaussian to the distribution (shown as a blue dash-dotted line) to the data, and the best fit parameters are given in the panel.  We will use this age distribution in constructing synthetic CMDs to test if other regions of the CMD are consistent with the estimated SFH.

We then generated two synthetic cluster CMDs.  The first is the expected distribution if the eMSTO was due to an age spread (i.e. adopting the age spread in Fig.~\ref{fig:ngc1806_age_dist_msto}).  The second was under the assumption of a single age population (i.e., an SSP) with an age of 1.44 Gyr.  Synthetic stellar populations were then made adopting a Salpeter~(1955) stellar IMF and the Marigo et al.~(2008) isochrones, applying representative errors taken from the observations.  Stars along the same colour/magnitude region used to define the SGB and the RC in the observations were selected and analysed in the same way as the observations.

In Appendix~\ref{sec:appendix_test} we test all the used techniques against synthetic clusters with known (input) age spreads.  Additionally, we test our relatively simplistic age spread estimate based on the cut across the eMSTO (note that this is the same technique adopted by Goudfrooij et al.~(2011b; 2014)) against a full SFH based on the {\sc STARFISH} package.

\subsubsection{The morphology of the Sub-Giant Branch (SGB)}

For the SGB, we can quantify the offset between the observed distribution and that expected from a population with an extended SFH or a single burst.  This is done by measuring the magnitude offset of each observed star (or a star from our synthetic CMD) identified as being part of the SGB from a nominal isochrone. 

\begin{figure}
\centering
\includegraphics[width=8cm]{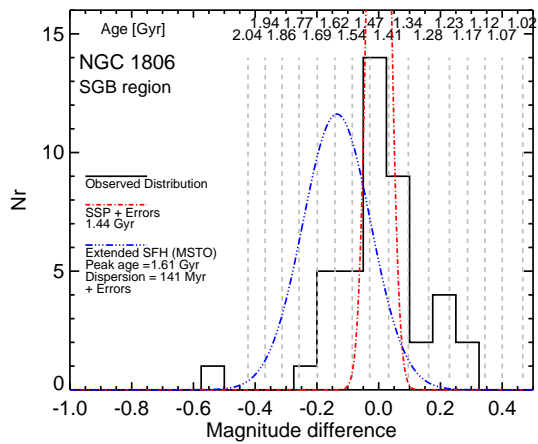}

\caption{Analysis of the SGB of NGC~1806.  The histogram shows the observed magnitude difference between each star identified as being part of the SGB and a nominal isochrone (chosen to be $1.44$~Gyr).  Vertical dashed lines show the expected magnitude difference of different isochrones, labelled in the panel.  The dash-dotted (red) line shows the expected distribution of stars if the underlying distribution was an SSP convolved with the observational errors.  The dash-dot-dot (blue) line shows the expected distribution adopting the parameters that best fit the eMSTO region (assuming that an age spread is the cause).  All distributions shown are for the same number of stars in the SGB region ($N = 41$ in this case).}
\label{fig:ngc1806_age_dist_sgb}
\end{figure}

We adopt an isochrone at 1.44~Gyr, as this best describes the observed SGB in this cluster,  and looked at the difference between each observed SGB star and the nominal isochrone.  The results are shown in Fig.~\ref{fig:ngc1806_age_dist_sgb}.  We also show the expected magnitude difference between other isochrones and the nominal isochrone as vertical dashed lines (labelled in the panel).  

As can be seen in Fig.~\ref{fig:ngc1806_age_dist_sgb}, the extended SFH simulation (the average of 1,000 monte carlo realisations) fails to reproduce both the peak of the observed distribution as well as the width.  The expected and observed peaks in the distribution are offset by $\sim0.2$~mag.  
The SSP simulation well describes the width of the core of the observed histogram, but the observations display extended tails on both sides, suggesting that photometric errors alone cannot explain the full distribution.  However, we note that the observed spread in the distribution (corrected for photometric uncertainties) is an upper limit to the actual spread, as we have not corrected for the effects of differential extinction, nor binarity, in our sample.  Additionally, if stellar rotation is affecting the MSTO then we may expect to see some influence on the SGB, which causes some spread in the observed SGB stars (e.g., Girardi et al.~2011).

Finally, we note that the base of the red-giant branch also appears narrower than would be expected if large age spreads were present.  Again, the stars cluster along the youngest isochrone that fits the MSTO region. 

\subsubsection{Red Clump Position}

\begin{figure}
\centering
\includegraphics[width=8cm]{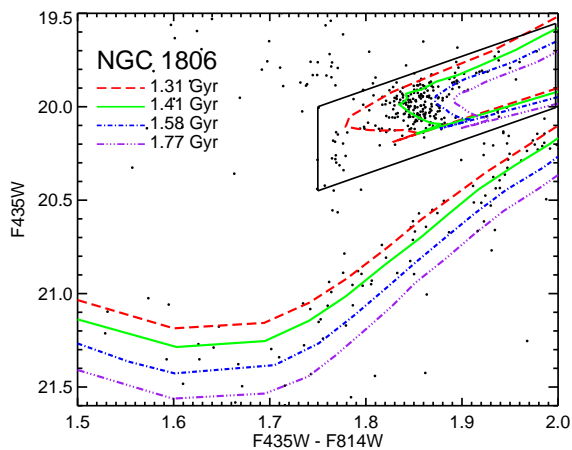}

\caption{The CMD of NGC~1806 highlighting the base of the red branch and red clump (RC).  Four isochrones are over plotted.  Note that ages younger than 1.32~Gyr or older than 1.58~Gyr do not fit the RC morphology.  The narrowness of the observed base of the RGB and RC location are inconsistent with significant age spreads within the cluster.  The box shows the region from which the star-formation history of RC stars was inferred.}
\label{fig:ngc1806_cmd_rc}
\end{figure}

In Fig.~\ref{fig:ngc1806_cmd_rc} we show a zoom in of the CMD of NGC~1806 centred on the base of the red giant branch and the red clump (RC).  Additionally, we show four isochrones with ages between 1.31 and 1.77~Gyr.  As was the case for the SGB, we note that the position of the observed stars do not conform to the expectations if there was a significant age spread present within the cluster.  While a significant spread in colour is expected, there is little spread in observations.  

\begin{figure}
\centering
\includegraphics[width=8cm]{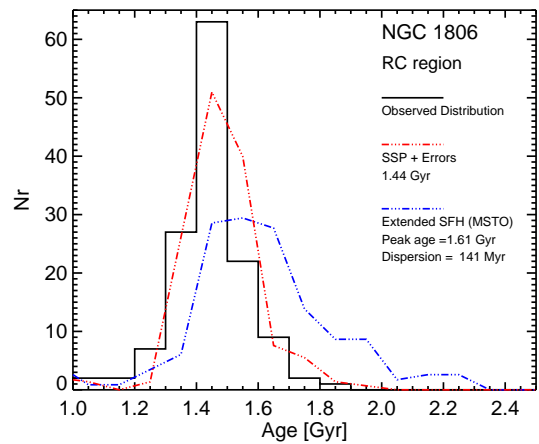}

\caption{The inferred star-formation history of NGC~1806 based on the morphology of the red clump (RC).  }
\label{fig:ngc1806_age_dist_rc}
\end{figure}

We quantify this in a similar way as was done for the SGB.  We simply assign an age to each of the observed (and synthetic) stars based on the closest isochrone.  The results are shown in Fig.~\ref{fig:ngc1806_age_dist_rc}.  Note that the RC width is consistent with the expectations of an SSP, while an extended SFH results in a distribution that is clearly at odds with the observations.  The reason why this ``pseudo-age distribution" is not symmetric (the input age distribution is) is that the isochrones bunch up at older ages at the RC, so small photometric errors may lead to large age differences.

\subsubsection{The Effects of Binarity}

We have explored the effect of binaries on the SGB morphology, by creating CMDs of synthetic clusters with and without binaries.  For the binary populations we assume a flat mass ratio and differing binary fractions, ranging from zero (only single stars) to 1 (all stars have binary companions).  Only in the cases of high binary fraction was the SGB significantly affected, shifting the stars to brighter magnitudes.  For realistic binary fractious we found that the SGB is only marginally broadened by binaries, consistent with the more detailed simulations of Milone et al.~(2009 - their Figs.~$31-36$). The reason for this is the relatively low fraction of near-equal mass binaries in the clusters ($10-15$\% - Milone et al.~2009).  Due to the intrinsic brightness of SGB stars, low mass companions do not significantly contribute to the combined flux of a binary pair. 

Due to the relatively short duration of the RC phase in stellar evolution, and the fact that RC stars are much brighter than their MS counterparts, binaries are not expected to have a large effect on the RC position or morphology.  Only in the rare circumstance that a RC star has an RC star binary companion would the magnitude be significantly affected.

\subsection{NGC~1846}

\begin{figure}
\centering
\includegraphics[width=8cm]{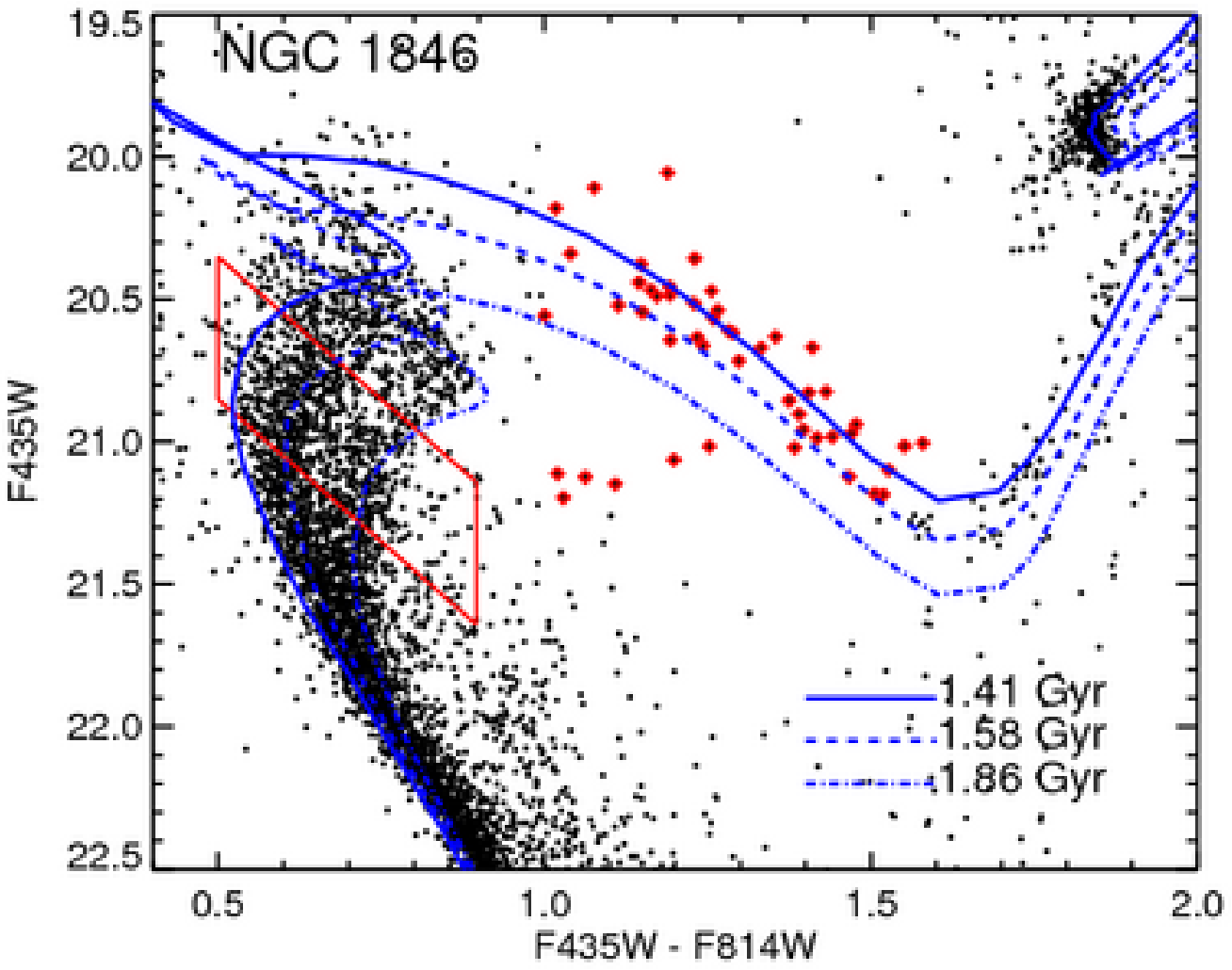}

\caption{The same a Fig.~\ref{fig:ngc1806_cmd} but now for NGC~1846.}
\label{fig:ngc1846_cmd}
\end{figure}

We have carried out a similar analysis of the colour-magnitude diagram of NGC~1846, again focussing on the SGB and RC regions.  The CMD of NGC~1846 is shown in Fig.~\ref{fig:ngc1846_cmd}.  As was the case for NGC~1806, different authors have suggested different age spreads for this cluster, with Milone et al.~(2009) suggesting a spread of $\sim250$~Myr, Rubele et al.~(2013) finding $330$~Myr and Goudfrooij et al.~(2011b) suggesting $\sim600$~Myr.  We adopt a nominal isochrone of $1.4$~Gyr for our analysis. 

\begin{figure}
\centering
\includegraphics[width=8cm]{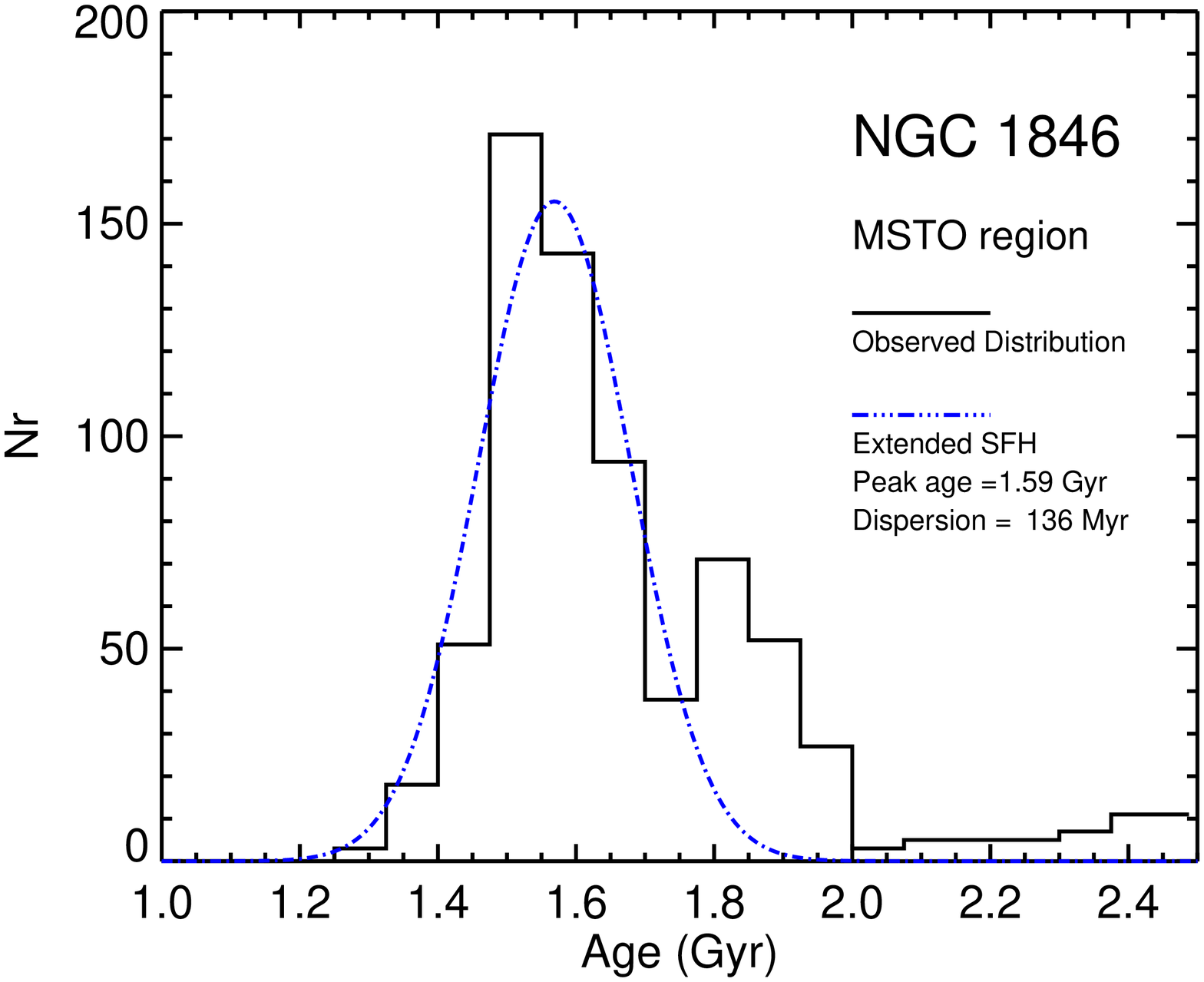}
\caption{The same as Fig.~\ref{fig:ngc1806_age_dist_msto} but now for NGC~1846.}
\label{fig:ngc1846_age_dist_msto}
\end{figure}

\begin{figure}
\centering
\includegraphics[width=8cm]{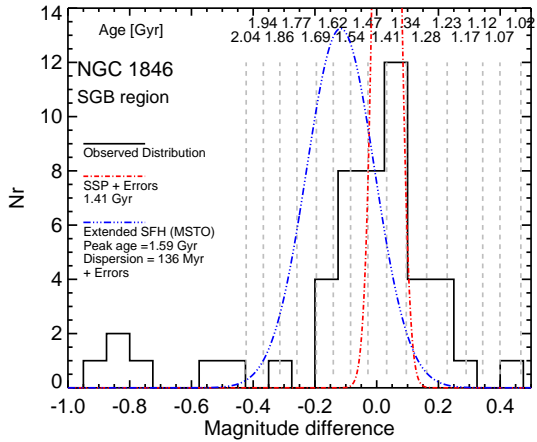}

\caption{Analysis of the SGB of NGC~1846 (similar to Fig.~\ref{fig:ngc1806_age_dist_sgb}).  }
\label{fig:ngc1846_age_dist_sgb}
\end{figure}

\begin{figure}
\centering
\includegraphics[width=8cm]{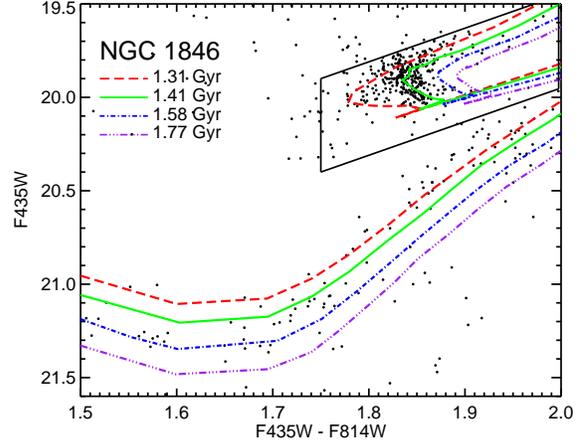}

\caption{The CMD of NGC~1846 highlighting the base of the red branch and red clump (RC), similar to Fig.~\ref{fig:ngc1806_cmd_rc}}
\label{fig:ngc1846_cmd_rc}
\end{figure}

\begin{figure}
\centering
\includegraphics[width=8cm]{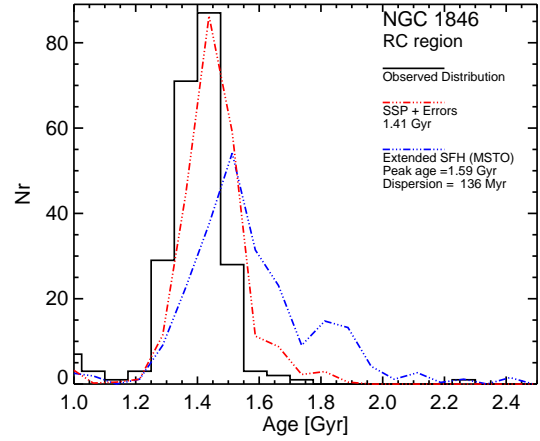}

\caption{The inferred star-formation history of NGC~1846 based on the morphology of the red clump (RC).  }
\label{fig:ngc1846_age_dist_rc}
\end{figure}

We carried out the same analysis as was done for NGC~1806, and the results are shown in Figs.~\ref{fig:ngc1846_age_dist_msto}, \ref{fig:ngc1846_age_dist_sgb}, \ref{fig:ngc1846_cmd_rc} \& \ref{fig:ngc1846_age_dist_rc}.

As was found for NGC~1806, the expected SGB morphology for an extended SFH (as inferred from the eMSTO region) is clearly offset from the observed distribution (Fig.~\ref{fig:ngc1846_age_dist_sgb}).  The SGB morphology suggests a significantly lower age ($\sim1.4$~Gyr) than that inferred from the eMSTO region ($\sim1.6$~Gyr), and also a narrower age spread.  The inferred age from the SGB is that of the youngest isochrone needed to match the eMSTO, which is $1.4$~Gyr, similar to that found for NGC~1806.  


From this we conclude that the SGB morphology is inconsistent with the presence of a significant age spread within the cluster.  Additionally, from Figs.~\ref{fig:ngc1846_cmd_rc} \& \ref{fig:ngc1846_age_dist_rc} it is clear that the RC position and morphology is also inconsistent with the inferred age spread from the MSTO, being best described by a single isochrone (an SSP).

As was the case for NGC~1806, we again note that we have not included the effects of binarity nor differential extinction (and also for the potential spread caused by stellar rotation), hence any observed spread is an upper limit to the intrinsic spread in the distribution.  Additionally, there is likely some amount of stellar field star contamination in the used CMD.

We have also tested our results using different filter combinations (e.g., including $F555W$) based on new photometry, Milone et al.~(in prep), and find consistent results.

Our results are not in agreement with those reported by Rubele et al.~(2013) who fit the SFH of NGC~1846 with the {\sc STARFISH} code (Harris \& Zaritsky~2001) and found that the CMD was best reproduced by an extended SFH.  Their fit included the SGB and RC portions of the CMD, as well as the MSTO, so it could have found an inconsistency between the different CMD regions.  The code, however, is weighted to regions with the most stars, hence was heavily weighted to the MSTO region of the CMD.  Figure~7 of Rubele et al.~(2013) clearly shows high $\chi^2$ values for this cluster at the position of the RC, inferring that the best fitting extended SFH did not reproduce this region, consistent with what we have reported here.  

Mackey et al.~(2013) have found evidence that NGC~1846 displays systematic rotation, similar in magnitude to the velocity dispersion of the cluster.  One explanation for this phenomenon is that a dynamically colder ``2nd generation" formed with a net rotation within an existing ``1st generation" (e.g., Bekki~2010).  However, since the cluster does not display any abundance variations (Mackey et al. in prep.) it is not possible to differentiate between the ``1st" and ``2nd generations" (only the ``2nd generation" stars would be expected to show rotation in this model).  Hence, it may simply be that the cluster as a whole rotates, with no relation to the eMSTO phenomenon or any potential age spreads within the cluster.  Indeed, rotation may be a common property of massive clusters, and has been observed in young massive clusters, e.g., R136 and Glimpse-C01, H{\'e}nault-Brunet et al.~(2012) and Davies et al.~(2011), respectively.


\section{Discussion and Conclusions}
\label{sec:conclusions}

The two clusters analysed in the present work (NGC~1806 \& NGC~1846) display sub-giant branches (SGBs) and red clumps (RCs) that are significantly narrower and offset from what would be expected if age spreads of $200-600$~Myr would be present within the clusters.  The observed spreads are in fact upper limits to the actual spreads, as we have not accounted for the affects of differential extinction (Platais et al.~2012) or binarity in our sample (or for the potential effects of stellar rotation).  Li et al.~(2014b) have found a similar behaviour in the CMD of NGC~1651, confirming that the lack of spreads in the SGB is common in clusters that display an eMSTO.  Hence, our results are not consistent with the interpretation of age spreads being the cause of the eMSTO phenomenon, in contradiction with previous claims (e.g., Goudfrooij et al.~2011b; Rubele et al.~2013; Correnti et al.~2014).   

Rubele et al.~(2013) fit the star-formation history of two intermediate age clusters, NGC~1846 and NGC~1783, under the assumption that the eMSTO was due to an age spread.  The authors used the {\sc STARFISH} star-formation history code (Harris \& Zaritsky~2001) which compares synthetic observations based on linear combinations of theoretical isochrones (plus errors) to the observations, performing a $\chi^2$ minimisation.  The authors report age spreads of $250-300$~Myr for the two clusters.  Much of the weight of the analysis comes from the MSTO region, due to the large number of stars in that phase, and a relatively low weight is applied to portions of the CMD with fewer stars (e.g., SGB, RGB, RC).  However, we note that for both clusters, the best fit extended SFH fails to reproduce the morphology of the red clump (in particular in the $F435W-F814W$ filters), in the sense that the models predict too many red RC stars.  We also note that the position of the RC was not matched in the analyses of NGC~1806, 1846, 411, 1718 and 2203 by Goudfrooij et al.~(2011a; 2014), again calling into question the interpretation of large spreads within these clusters.  This is similar to the results found in the present work.  We will analyse the CMD of NGC~1783 in detail in a future work (Niederhofer et al.~in preparation). 


Based on the SGB and RC locations (and also on the base of the red-giant branch), we infer that NGC~1806 and 1846 have younger ages than has been reported in the past, with the best fitting age from the SGB agreeing with the youngest age inferred from the MSTO region (see \S~\ref{sec:appendix_metallicity} for a discussion of the degeneracies between age and metallicity in the CMD fitting of these clusters).  Hence, it appears that some effect is causing stars to move red-ward in colour space at the MSTO.  NGC~1651 displays the same behaviour, with an inferred age of $\sim2$~Gyr, suggesting that it is a general property of the eMSTO intermediate age clusters, and not an age effect (Li et al.~2014b).  
This is the behaviour predicted if stellar rotation was the cause of the eMSTO (Bastian \& de Mink~2009; Yang et al.~2013).  At present, it is unclear how rotation will affect the SGB and RC, which we will leave to a future work.

Goudfrooij et al.~(2014) found a correlation between the fraction of blue/bright stars on the MSTO (i.e., young stars if interpreted as due to age effects) and the fraction of ``secondary RC" stars to the main body of RC stars, i.e.  stars that are bluer and fainter than the main red clump.  The authors use this to argue for the presence of age spreads within their cluster sample, although, as shown here, the main body of the RC does not show evidence for age spreads.


Goudfrooij et al.~(2011, 2014) have suggested that a correlation exists between the width of the eMSTO and the escape velocity of the cluster.  However, this result depends heavily on large ``correction factors" to change the current observed escape velocities to the ``initial ones".  Without these correction factors many of the clusters that display the eMSTO feature would have lower escape velocities than young clusters that have been shown not to have age spreads within them (e.g., Bastian \& Silva-Villa~2013; Niederhofer et al.~2015).  The applied ``correction factors" are based on a number of extreme assumptions.  First, the models of cluster dissolution that are adopted are not appropriate to clusters in the LMC (e.g., the tidal field experienced by the model clusters is a Milky Way like Galaxy at $4$~kpc from the galactic centre, unlike the much lower tidal field of the LMC - e.g., Lamers et al.~2005).  Secondly, it is assumed that the clusters begin their lives tidally limited in order to maximise cluster mass-loss, whereas most clusters in the LMC/SMC that have been studied to date are not tidally limited (e.g., Glatt et al.~2011).  This means that cluster mass loss is in fact minimal, regardless of the initial state of mass segregation present in the cluster.  Hence, realistic correction factors would be much smaller than those adopted in Goudfrooij et al.~(2011b; 2014) and Correnti et al.~(2014). 

We can also use the cluster population properties of the LMC to test the Goudfrooij et al. (2011b; 2014) scenario. While the current masses of the eMSTO clusters are similar to the most massive YMCs (e.g., NGC~1850, 1866 and 1856 - Niederhofer et al.~2015), Goudrooij et al.~(2011; 2014) posit that the intermediate age clusters were much more massive (by a factor of $\sim4$) at birth (and also significantly denser).  Based on size-of-sample effects (e.g., Gieles \& Bastian~2008) we would expect that the SFR of the LMC was $\sim4$ times higher while the eMSTO clusters were forming (c.f., Maschberger \& Kroupa~2011).  However, no such increase in the SFH, based on resolved stars, is seen from $1-2$~Gyr ago relative to today in the LMC (e.g., Harris \& Zaritsky~2009; Weisz et al.~2013).  The lack of increase in the SFH from $1-2$~Gyr is consistent with the fact that the clusters in this age range would have been born with similar masses (at the upper end of the mass function) as that observed in the YMCs in the LMC today.  As discussed above, the clusters are unlikely to be tidally limited so their current masses should be good approximations of the initial masses.  This can be quantified by using the LMC cluster sample of Baumgardt et al.~(2013).  There are nine clusters more massive than $5 \times10^4$~\msun\ (a high value was taken to avoid any completeness limit issues) with ages less than $500$~Myr, or 0.018 clusters/Myr.  Taking the same mass limit but looking at the $1-2$~Gyr age range, we find 15 such clusters, or 0.015 clusters/Myr, very similar to that found for the past $500$~Myr.  Hence, the cluster population suggests that the cluster formation rate, like the SFR derived from CMD analyses, was roughly the same for the past $500$~Myr as it was from $1-2$~Gyr in the LMC.  Neither the SFH, based on resolved stars, nor the cluster population suggest an increase of a factor of $\sim4$ in the LMC from $1-2$~Gyr relative to now, that would be expected in the Goudfrooij et al.~(2011b; 2014) scenario. We conclude that it is unlikely that the eMSTO cluster had masses, hence escape velocities, significantly higher in the past. 

Girardi et al.~(2009) studied at SMC intermediate age clusters, NGC~419, focussing on the RC morphology.  They find that the MSTO is broader than would be expected from a nominal isochrone, and suggest an age spread of $\sim500$~Myr within the cluster.  We note that the SGB appears to show the same morphology, i.e. clusters towards the youngest isochrones, as NGC~1806 and NGC~1846 studied here.  However, the RC position is consistent with that expected for an older age, unlike that found for the clusters studied here.

Our results corroborate a number of other results from the literature, that have shown that age spreads of hundreds of Myr are not present within clusters.  As discussed in the introduction, a number of studies have searched for evidence of ongoing star-formation, or large age spreads, within older ($>10$~Myr) clusters, with none found (e.g., Perina et al.~2009; Larsen et al.~2011; Bastian \& Silva-Villa~2013, Bastian et al.~2013a; Niederhofer et al.~2015).  Additionally, Cabrera-Ziri et al.~(2014) used an integrated optical spectrum of Cluster~1 in NGC~34, a $\sim10^7$\msun, 100~Myr cluster, to estimate its star-formation history.  The cluster was found to be well described by an SSP (i.e. no age spread was found), with upper limits of $10-20$\% of the cluster mass being made up of younger stars.  The authors have extended this study by looking at a number of other massive $10^6 - 10^8$~\msun\ clusters, including UV and optical photometry in the analysis, and have not found evidence for extended SFHs (Cabrera-Ziri et al.~in preparation).

Additionally, Bastian \& Strader~(2014) searched for gas and dust in 12 LMC/SMC YMCs, the presence of which is required if new stars are to form, and found none in any cluster, down to limits of $<1$\% of the stellar mass.  Cabrera-Ziri et al.~(2015) extended this analysis to study seven massive clusters ($>10^6$\msun) in the Antennae merging galaxies, using the Atacama Large Millimetre Array (ALMA) to search for gas within the clusters (with ages of $50-250$~Myr).  No gas was detected within the clusters, with an upper limit of $\sim8$\% of the stellar mass present in gas.  

Mucciarelli et al.~(2008) searched for chemical spreads (expected if large age spreads exist, in particular if self-enrichment scenarios are invoked) in three LMC clusters that display the eMSTO phenomenon, and no abundance spreads were found.  Mucciarelli et al.~(2014) have extended this analysis to NGC~1806 (also studied in the present work) and again the authors have not found any abundance spreads.  Finally, Mackey et al.~(in preparation) have also searched for abundance spreads, targeting NGC~1846, and like for the other young and intermediate age clusters studied to date, no chemical spreads were found.  While a lack of abundance spreads does not necessarily mean that age spreads cannot exist within the clusters, it does however require either 1) cluster/cluster mergers with different ages (although this would not lead to an extended MSTO, but rather a dual MSTO) or 2) that the cluster was able to remove stellar ejecta (e.g., the ejecta of AGB stars) at the same time as it was accreting primordial gas from its surroundings, and not have the two gas flows interact and mix.  Hence, significant fine-tuning would be required.

Taken together, we conclude that age spreads are not the cause of the eMSTO feature found in many intermediate age ($1-2$~Gyr) clusters.  Hence, alternative models, such as stellar rotation or interacting binaries should be investigated further.  New alternative scenarios should also be encouraged and developed.


\section*{Acknowledgments}

We are grateful to Antonino Milone for providing his photometry and catalogues.  We would like to thank Dougal Mackey, Gary Da Costa Selma de Mink, and Richard de Grijs for helpful discussions and comments on the manuscript.  The anonymous referee is thanked for suggestions that helped improve the paper.  NB is partially funded by a Royal Society University Research Fellowship. The results presented here are partially based on observations made with the NASA/ESA Hubble Space Telescope, and obtained from the Hubble Legacy Archive, which is a collaboration between the Space Telescope Science Institute (STScI/NASA), the Space Telescope European Coordinating Facility (ST-ECF/ESA) and the Canadian Astronomy Data Centre (CADC/NRC/CSA).

\appendix
\section{Testing the Adopted Methods}
\label{sec:appendix_test}

In order to test whether the age spreads derived from the eMSTO accurately reproduce the observations (i.e. whether the cut perpendicular to the MSTO is not affected by differing lifetimes of stars in that phase in the narrow mass range considered) we have performed a series of tests, which are discussed here.  We note, again, that this technique has been used previously, namely in Goudfrooij et al.~(2011a,b; 2014).

\subsection{Synthetic Distributions}
\label{sec:appendix_synthetic}

First we created a synthetic population of stars with an age spread similar to the ones inferred from the perpendicular cut across the eMSTO for NGC~1806 and NGC~1846.  For this we adopted the same isochrones used to analyse the observations (same metallicity), assumed a Salpeter~(1955) stellar initial mass function, and created a population with a Gaussian age spread with a peak at 1.59~Gyr and a dispersion of 136~Myr.  No extinction was added to the data.  Additionally, we adopted the same distance modulus used for NGC~1806.

\begin{figure}
\centering
\includegraphics[width=8cm]{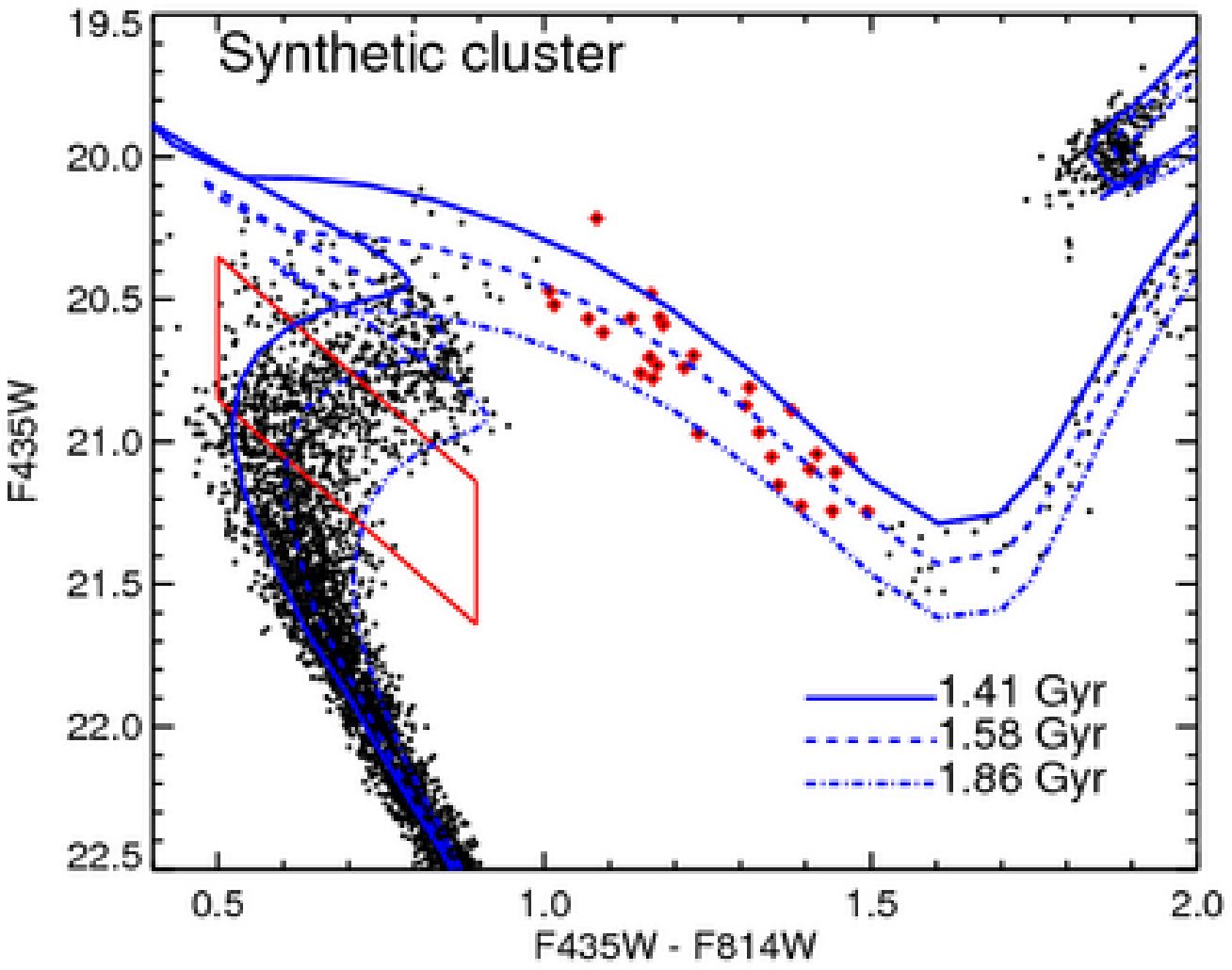}

\caption{The same as Fig.~\ref{fig:ngc1806_cmd} but now for the synthetic population.  The synthetic cluster has a gaussian SFH, with a peak at 1.59~Gyr and a dispersion of 136~Myr.}
\label{fig:syn_cmd}
\end{figure}

\begin{figure}
\centering
\includegraphics[width=8cm]{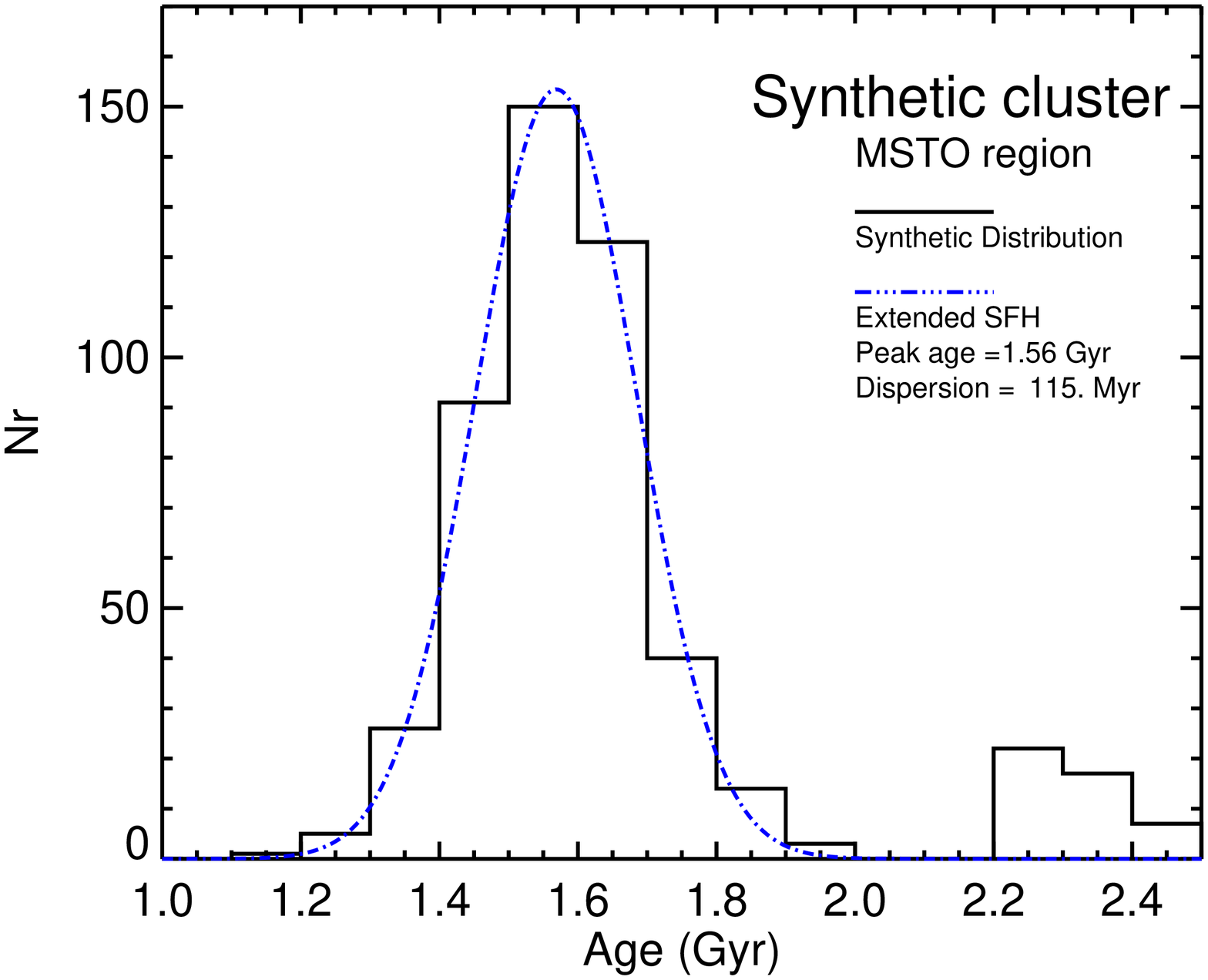}

\caption{The same as Fig.~\ref{fig:ngc1806_age_dist_msto} but now for the synthetic population.}
\label{fig:syn_age_dist_msto}
\end{figure}

\begin{figure}
\centering
\includegraphics[width=8cm]{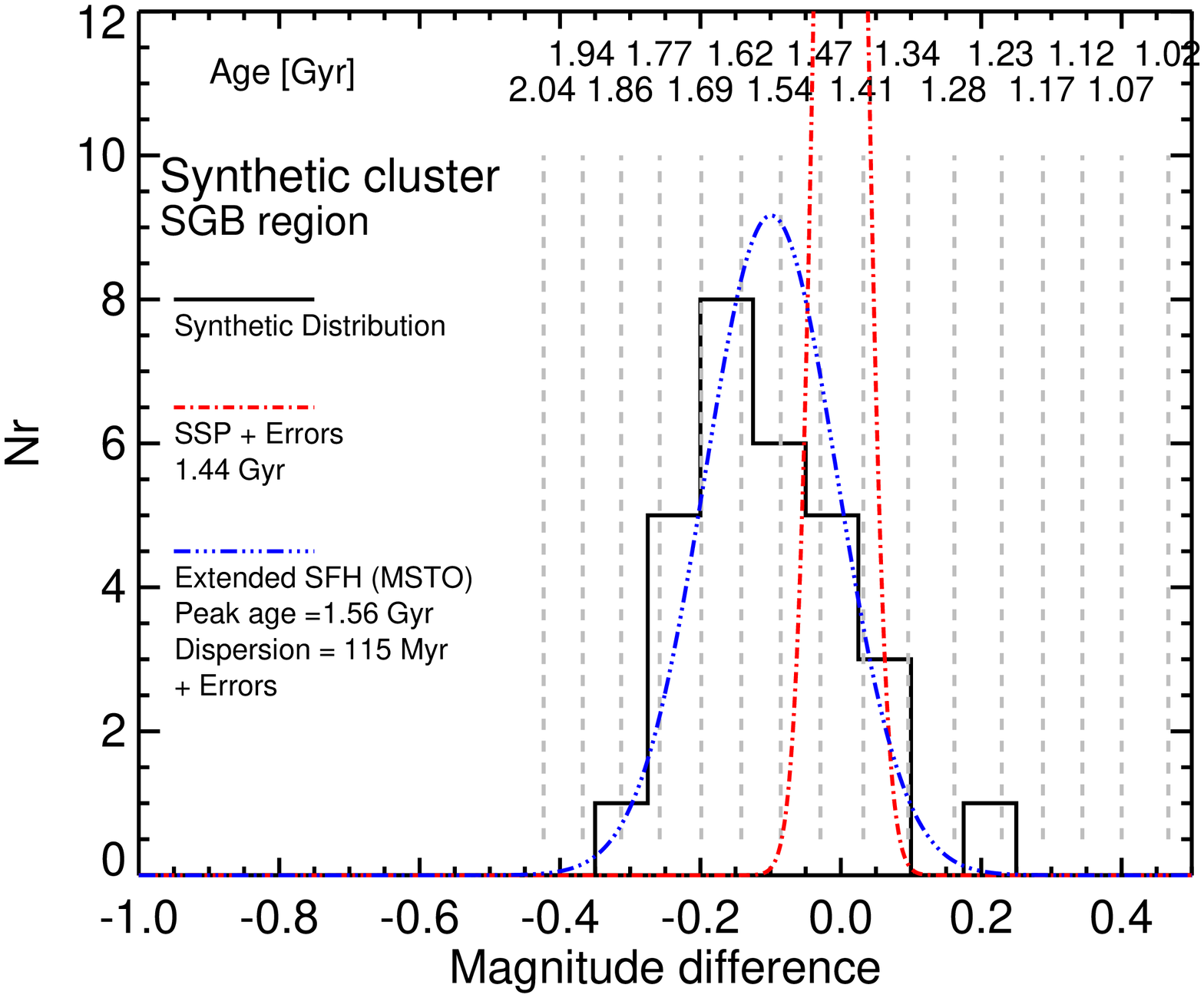}

\caption{The same as Fig.~\ref{fig:ngc1806_age_dist_sgb} but now for the synthetic population.}
\label{fig:syn_age_dist_sgb}
\end{figure}

\begin{figure}
\centering
\includegraphics[width=8cm]{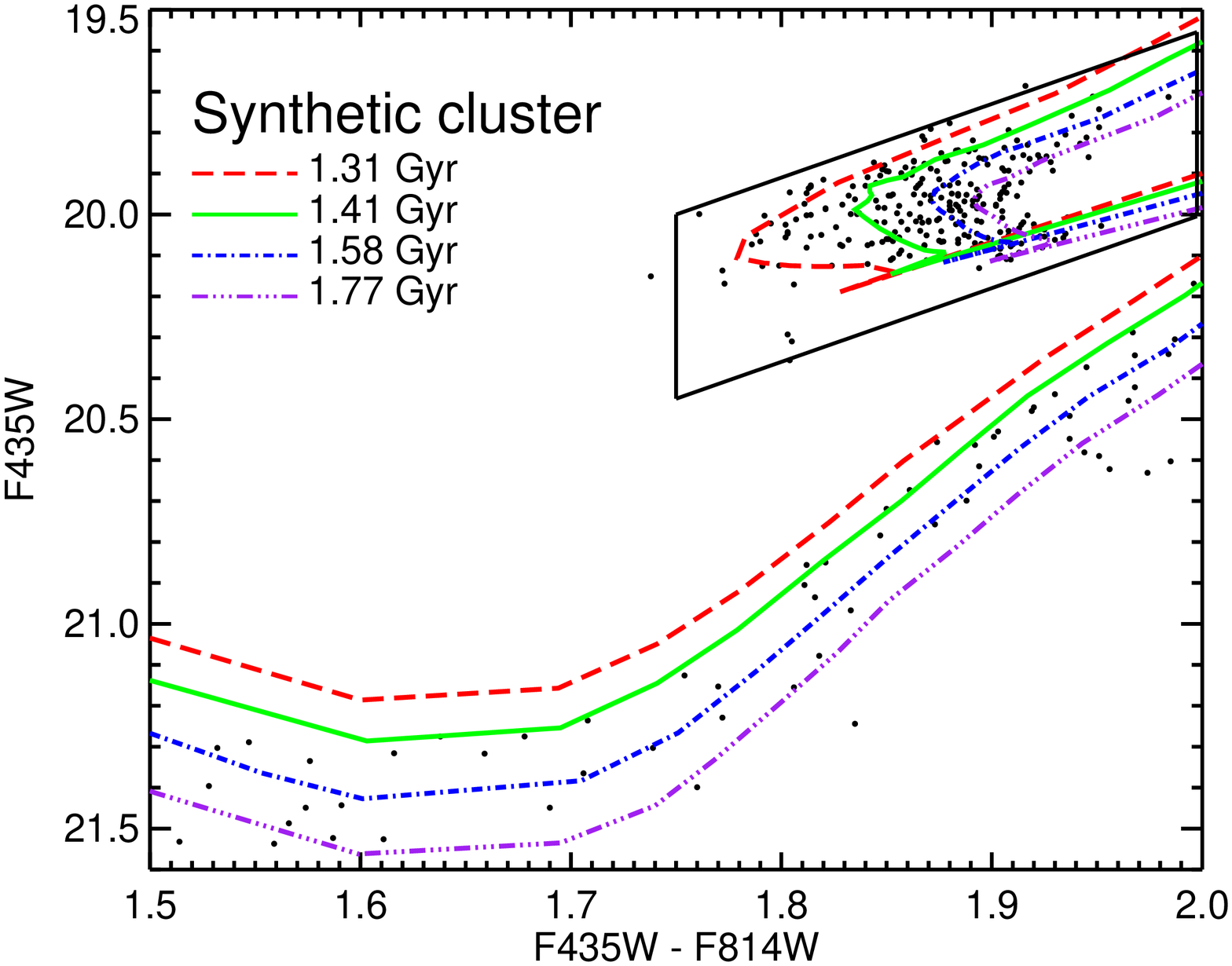}

\caption{The same as Fig.~\ref{fig:ngc1806_cmd_rc} but now for the synthetic population.}
\label{fig:syn_cmd_rc}
\end{figure}

\begin{figure}
\centering
\includegraphics[width=8cm]{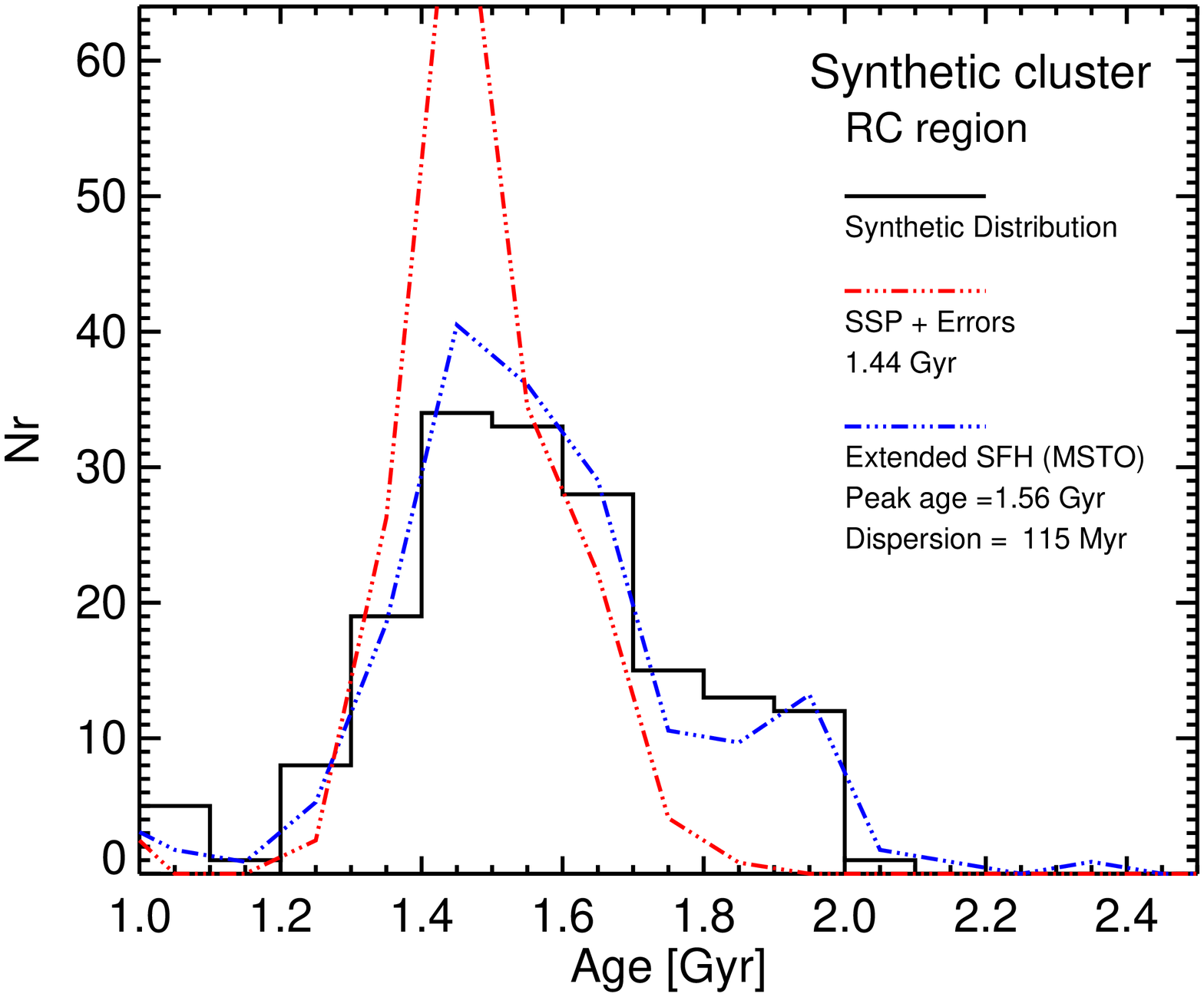}
\caption{The same as Fig.~\ref{fig:ngc1806_cmd} but now for the synthetic population. }
\label{fig:syn_age_dist_rc}
\end{figure}

We then ran this synthetic distribution through the same procedures used in the present work.  The results are shown in Figs.~\ref{fig:syn_cmd}$-$\ref{fig:syn_age_dist_rc}.  We find that the peak of the age distribution is well recovered, and that the dispersion found in the analysis techniques results in a slightly ($\sim10-20$\%) smaller dispersion than is input.  Hence, we are slightly underestimating the actual age spread within the cluster using the analyses technique of cutting across the eMSTO.  However, as can be seen in Figs.~\ref{fig:syn_cmd}$-$\ref{fig:syn_age_dist_rc}, such an age spread would easy be detected in the morphologies and positions of the SGB and RCs.

We have run the above tests for a number of ages between 1 and 2 Gyr and age dispersion of $100-150$~Myr and found consistent results.  However, for ages of $\sim1.8$~Gyr and above it is difficult to infer short age dispersions using the techniques used here.  Age spreads can still be recovered, however their actual extent becomes highly uncertain.  This is discussed in detail in Keller, Mackey \& Da Costa~(2011).

\subsection{{\sc STARFISH} SFH Fitting}
\label{sec:appendix_fitting}

In addition to the synthetic cluster experiments discussed above, we have also fit the observed CMDs of NGC~1806 and NGC~1846 with the {\sc STARFISH} (Harris \& Zaritsky~2001) star-formation history code.  This is the same technique which was used by Rubele et al.~(2013).  We adopted the same isochrones as used in the other analyses performed in the current work; adopting the same metallicity, distance modulus, and extinction as before.

This {\sc STARFISH} code finds the best-fitting model by creating linearly combined synthetic CMDs of simple stellar populations (i.e. single isochrones) and comparing them to the data. We used isochrones between 8.70 $\le$ log(t/yr) $\le$ 9.48, logarithmically spaced in steps of 0.03. As with the previous experiments, we adopt a Salpeter~(1955) stellar IMF.
We created an analytical model for the observational errors, assuming an error for the brightest stars in the clusters of about 0.01 which increases exponentially to 0.04 at a magnitude of about 24 in the {\it F435W} band. 
Furthermore, we assumed a binary fraction of 0.3 for both clusters.  In order to compare directly with the adopted SFH based on the cut along the MSTO, we restricted the {\sc STARFISH} fitting to a region around the MSTO. We chose boxes with the following limits: 0.5 $\le F435W-F814W \le$ 1.5 and 19.5 $\le F435W \le$ 22.00.

\begin{figure}
\centering
\includegraphics[width=8cm]{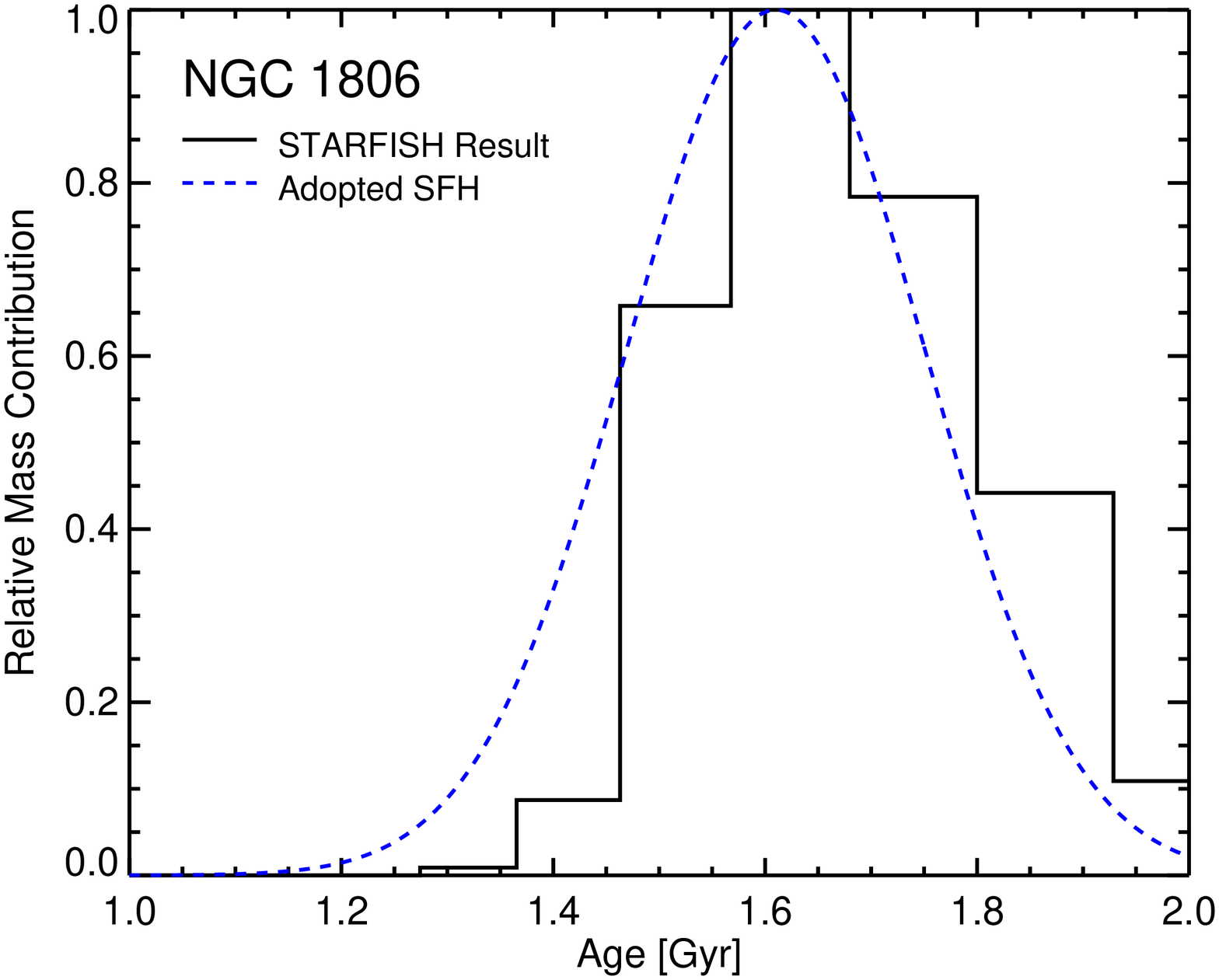}
\includegraphics[width=8cm]{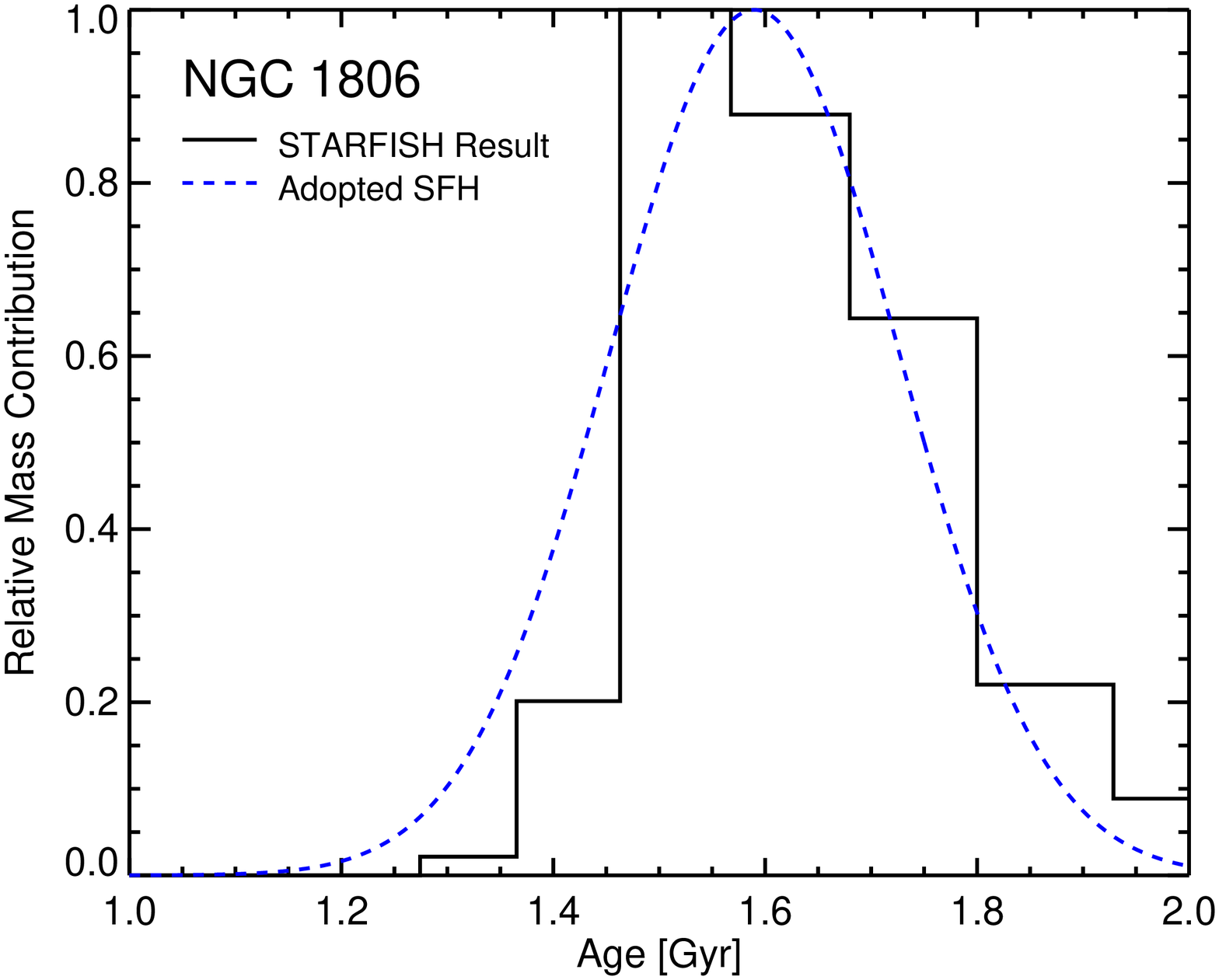}
\caption{A comparison between the estimated star-formation history based on the observed eMSTO in NGC~1806 (top panel) and NGC~1846 (bottom panel).  Two methods have been used and are shown in each of the panels.  The dashed (blue) lines represent the adopted SFH based on a cut perpendicular to the MSTO (see \S~\ref{sec:msto_cut}).  The histogram shows the results of fitting the MSTO region of each cluster with the star-formation history fitting code {\sc STARFISH}.  Note that both methods result in very similar age distributions.}
\label{fig:starfish}
\end{figure}

The results for NGC~1806 and NGC~1846 are shown in Figs.~\ref{fig:starfish}.  The histogram shows the derived SFH from {\sc STARFISH} while the dashed (blue) lines show the adopted distribution based on the cut across the eMSTO.  We find good agreement between the two methods, hence the analysis performed in the current work is able to provide a reasonable estimate of the potential age spread along the MSTO.  It is this potential age spread that has been tested against the morphology and position of the SGB and RC.  In both regions we do not find evidence for such age spreads.

\subsection{Metallicity Determination}
\label{sec:appendix_metallicity}

\begin{figure}
\centering
\includegraphics[width=8cm]{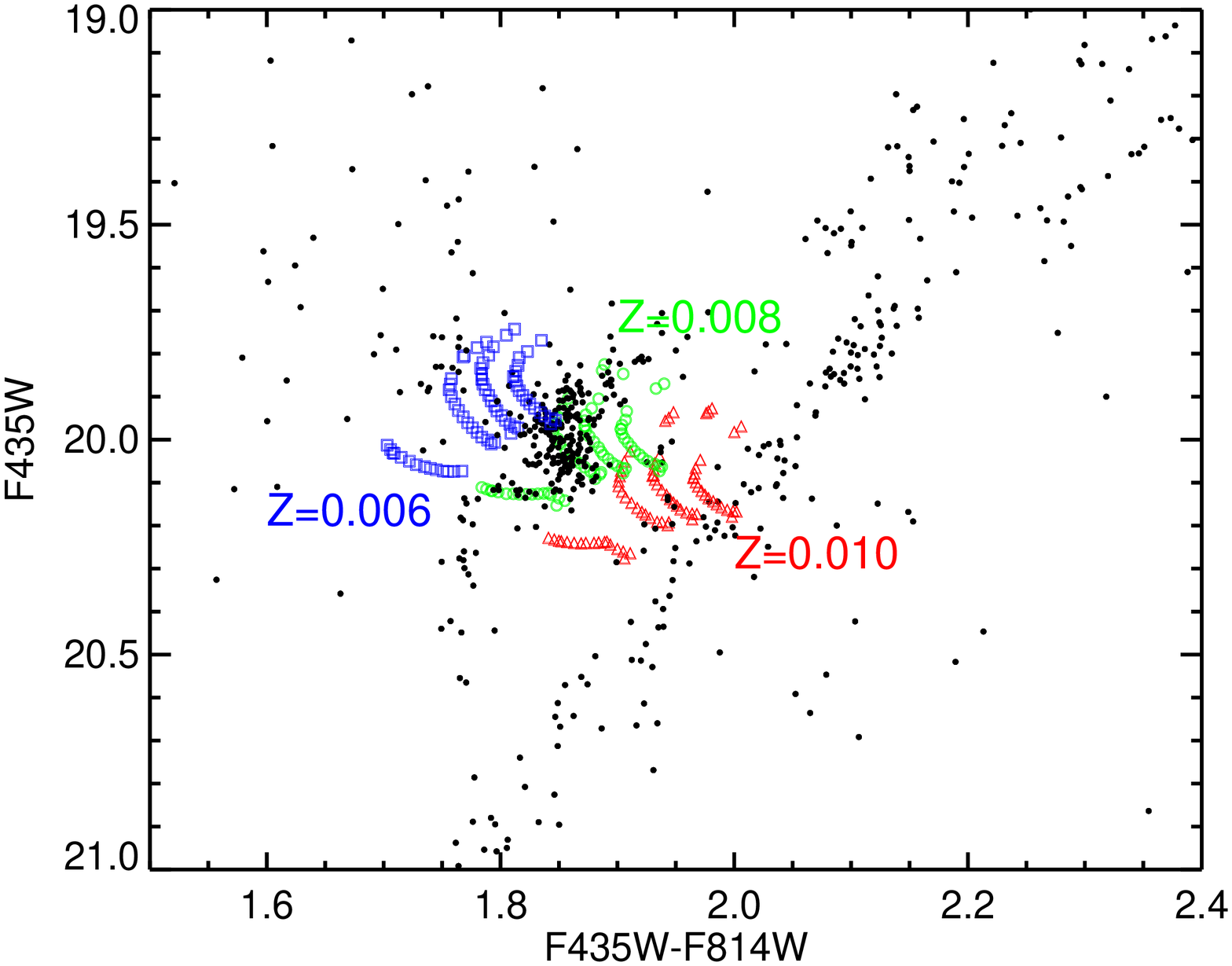}
\caption{The position of the RC as a function of metallicity ($Z=0.006$ shown as blue squares, $Z=0.008$ shown as green circles, $Z=0.010$ shown as red triangles) and age from the Marigo et al.~(2008) isochrones.  The ages shown are 1.2, 1.32, 1.44, 1.58 and 1.90~Gyr from left to right.   We have adopted the same distance modulus and extinction for NGC~1806 as in the other analyses of this cluster (\S~\ref{sec:obs} \& \S~{sec:ngc1806}). The filled points are the data for NGC~1806.}
\label{fig:metallicity}
\end{figure}

In order to test whether other combinations of age and metallicity can successfully reproduce the observed cluster CMDs, in Fig.~\ref{fig:metallicity} we show isochrones (Marigo et al.~2008) of various ages (1.32, 1.44, 1.58 and 1.9~Gyr) and metallicities ($Z=0.006, Z=0.008, Z=0.010$) focussing on the position of the red clump.  Additionally, we show the data of NGC~1806.  We adopt the same distance modulus and extinction for NGC~1806 as used in the text (\S~\ref{sec:obs} \& \ref{sec:ngc1806}).  The metallicity range shown is the same as the uncertainties for this cluster (and all clusters) in the sample of Goudfrooij et al.~(2011a,b, 2014) based on their analysis of the CMDs.  As can be seen, no combination of age and metallicity can fit the position of the RC for metallicities of $Z=0.006$ and $Z=0.010$ (for the adopted values of extinction and distance modulus).    

\begin{figure}
\centering
\includegraphics[width=8cm]{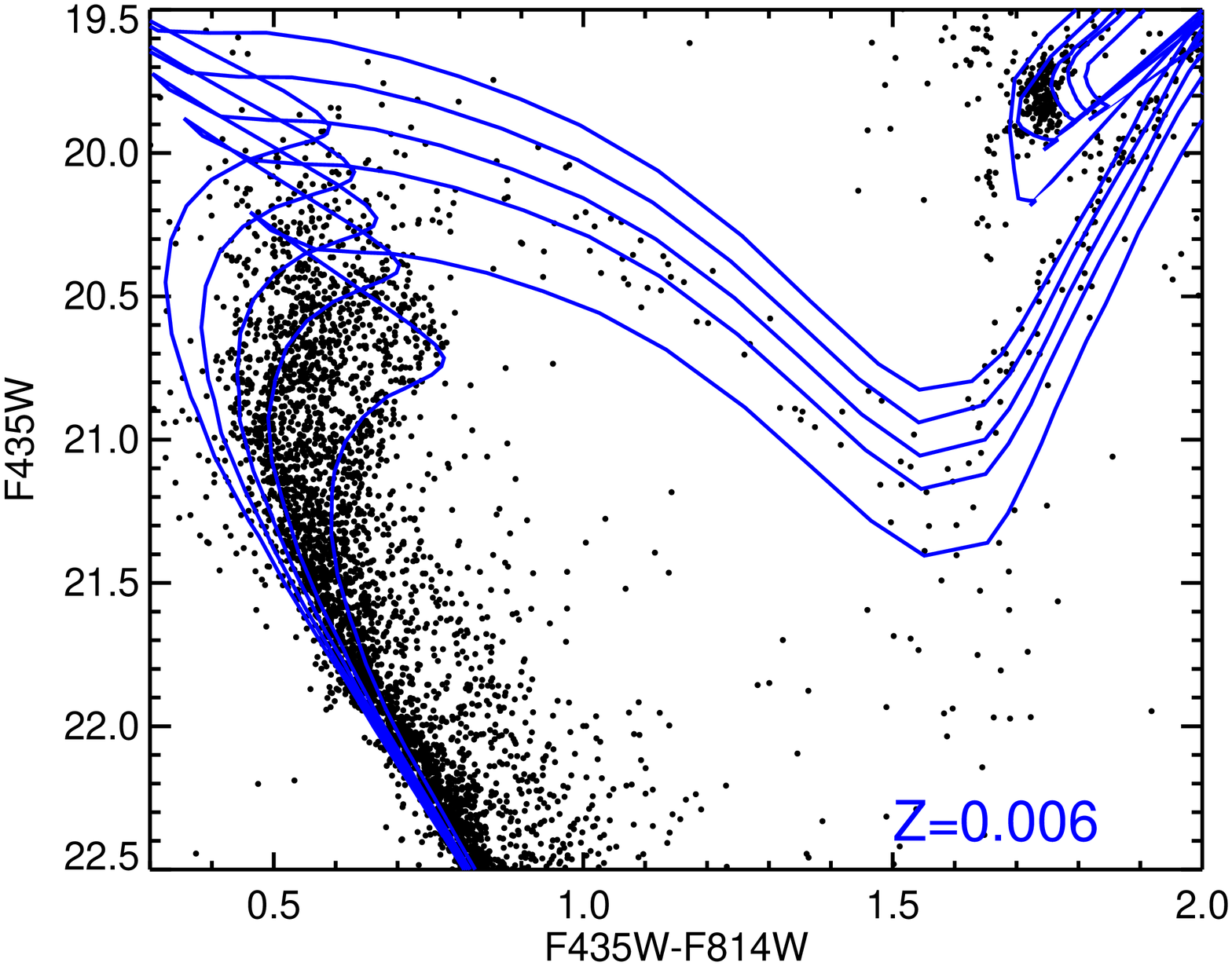}
\caption{The CMD of NGC 1806 showing the best fitting isochrones assuming $Z=0.006$, and adjusting the distance modulus and extinction in order to best reproduce the position of the RC.  The ages shown are 1.2, 1.32, 1.44, 1.58 and 1.90~Gyr from left to right (or top to bottom on the SGB).  The best fit corresponds to an $A_V$ and distance moduls of $0.2$ and $18.38$, respectively. }
\label{fig:metallicity2}
\end{figure}

\begin{figure}
\centering
\includegraphics[width=8cm]{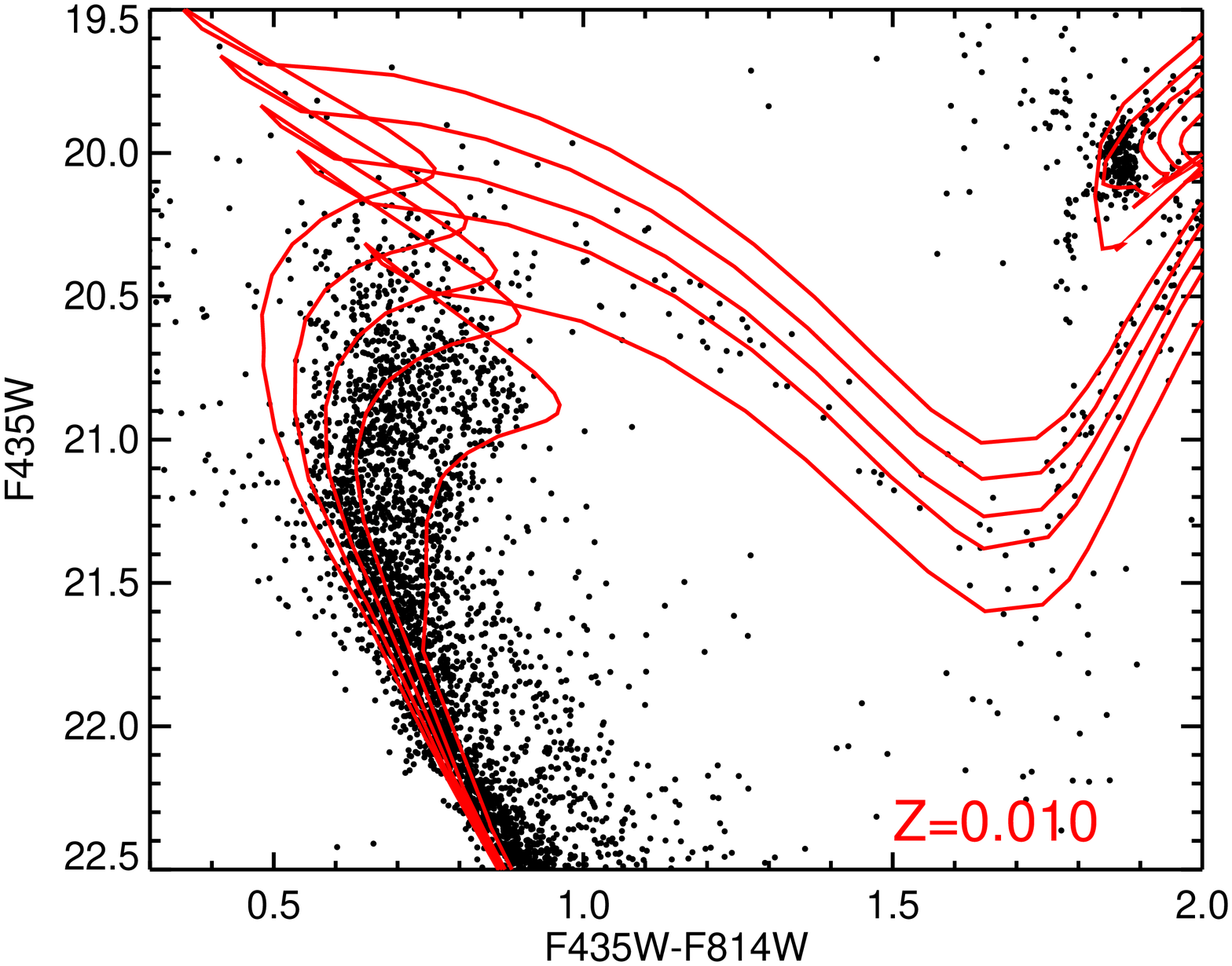}
\caption{The same as for Fig.~\ref{fig:metallicity2} but now showing the results for isochrones of $Z=0.010$.  The best fitting parameters are a distance modulus of 18.38~mag and an extinction of $A_V=0.03$~mag. The ages shown are 1.2, 1.32, 1.44, 1.58 and 1.90~Gyrfrom left to right (or top to bottom on the SGB). However, if this metallicity is adopted the RC is still best reproduced by the youngest isochrone needed to explain the eMSTO.}
\label{fig:metallicity3}
\end{figure}

We have also investigated whether such metallicities can be consistent with the data if we also allow the distance modulus and extinction to vary.  Based on the compact morphology of the RC, the age of the population is constrained to $>1.25$~Gyr for the metallicities considered here. In each case we adjusted the extinction and distance modulus to best match the position of the RC and the main sequence, for ages between $1.25$~Gyr and $2$~Gyr.  The results are shown in Figs.~\ref{fig:metallicity2} and \ref{fig:metallicity3}, for isochrones with $Z=0.006$ and $Z=0.010$, respectively.  For the case of $Z=0.006$, the best fitting parameters are $A_V=0.2$ and distance modulus$=18.38$.  This distance modulus would put NGC~1806 $\sim3.5$~kpc in front of the main body of the LMC.  For this choice of metallicity, we see that the RC is best fit with an age of $\sim1.4$~Gyr, which is the same as the blue side of the eMSTO.  This is the same as found (in \S~\ref{sec:conclusions}) for the case of $Z=0.008$.  However, in the $Z=0.006$ case the SGB is best described by an isochrone of age of $\sim1.5$~Gyr, slightly older than that found for the blue side of the eMSTO and position of the RC.

For the $Z=0.010$ isochrones, the best fit is obtained for an $A_{V}$ and distance modulus of 0.03 and 18.38, However we note a mis-match between the isochrones and the observed main sequence at faint magnitudes, which increases for fainter stars.  However, if the $Z=0.010$ isochrones are adopted, the position of the RC is still best reproduced by the youngest isochrone needed to explain the eMSTO ($\sim1.38$~Gyr).  In this case, the SGB is also consistent with a single isochrone, but one that slightly older than that needed to explain the RC ($\sim1.54$~Gyr for the SGB compared to $1.38$~Gyr for the RC and bluest part of the eMSTO).


We conclude that the ages derived for the clusters in this work based on the position and morphology of the SGB and RC do not suffer from a metallicity degeneracy, and that the SGB and RC are consistent with a single isochrone for all metallicities considered.  For metallicities slightly higher or lower than that adopted, the position of the RC still is best fit by the isochrone that best reproduces the blue side of the eMSTO.  However, in both cases ($Z=0.006$ and $Z=0.010$) the derived age based on the SGB is slightly older than that of the blue side of the eMSTO and the RC. Regardless of the adopted metallicity, the observed positions and morphologies of the SGB and RC are inconsistent with the large age spreads inferred from the eMSTO.

\bsp
\label{lastpage}
\end{document}